\documentclass[prl,aps,twocolumn]{revtex4}
\usepackage{graphicx}
\usepackage[fleqn]{amsmath}
\usepackage{amssymb}
\usepackage[varg]{txfonts}
\usepackage{color}

\def\figref#1{\mbox{Fig.~\ref{#1}}}
\def\eqref#1{\mbox{Eq.(\ref{#1})}}
\def\eqrefs#1{\mbox{Eqs.(\ref{#1})}}

\def\dd{{\rm d}}
\def\pp{{\partial}}
\def\sub#1{_{\rm #1}}

\begin{document}

\title{Active and reactive power in stochastic resonance for energy harvesting}

\author{Madoka Kubota} \email{kubota@dove.kuee.kyoto-u.ac.jp} 
\affiliation{Department of Electrical Engineering, Kyoto University, Katsura, Nishikyo, Kyoto, 615-8510 Japan}
\author{Ryo Takahashi} 
\affiliation{Department of Electrical Engineering, Kyoto University, Katsura, Nishikyo, Kyoto, 615-8510 Japan}
\author{Takashi Hikihara} 
\affiliation{Department of Electrical Engineering, Kyoto University, Katsura, Nishikyo, Kyoto, 615-8510 Japan}
\begin{abstract}
A power allocation to active and reactive power in stochastic resonance is discussed for energy harvesting from mechanical noise.
It is confirmed that active power can be increased at stochastic resonance, in the same way of the relationship between energy and phase at an appropriate setting in resonance.
\end{abstract}

\maketitle

\section{Introduction}
Since noise appears everywhere in our surroundings, the energy conversion from noise to controlled motion 
is a key to come it to use.
As an energy harvester from thermal noise, the molecular sized brownian ratchet was suggested, 
but this machine was verified its impossibility \cite{FeynmanI,TRKelly1999unidirectional}.
On the other hand, stochastic resonance (SR) has been suggested as a method for energy harvesting from noise \cite{CRMcinnes2008enhanced,KNakano2014feasibility,YZhang2014feasibility}. 
SR is a phenomena in which noise with moderate strength enhances SNR (Signal Noise Ratio) \cite{RBenzi1981mechanism,CNicoGNico1981stochastic}.
SR is possibly to apply in biological sensory organs \cite{JKDouglass1993noise,DFRussell1999use}, and the applications have expanded in medical use \cite{YKurita2013wear}, brain image enhancement \cite{VPSRallabandi2010magnetic}, and nano sized transistor \cite{oya2006stochastic,kasai2011control}. 

Here, we discuss SR for harvesting energy from white noise and the method to enhance the energy. 
This letter develops a concept of power factor correction as same as electrical systems, 
and surveys a power allocation to active and reactive power. 
Finally it is clarified that active power of controlled motion is maximized at SR.

\section{System and power equations}
\begin{figure}[tb]\centering
 \includegraphics[width=84mm]{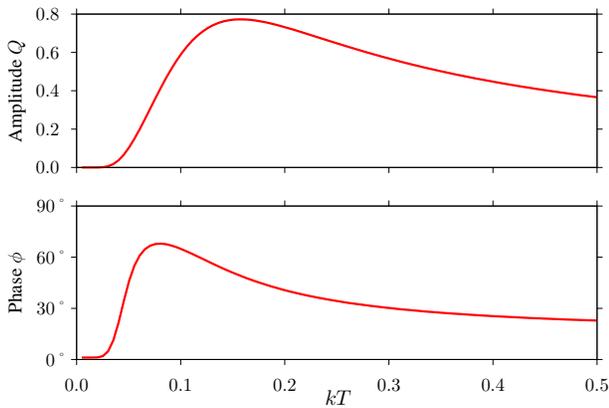}
 \caption{Behaviors of amplitude $Q$ and phase $\phi$ as functions of $kT$, where $m=0.02$, $\gamma m = 3.00$, $h=0.20$, and $\Omega = 0.04$. Both $Q$ and $\phi$ show their single peaks.}
 \label{fig:kT-QPhi}
\end{figure}
Suggested energy harvesters by the method of SR possess bi-stable potential \cite{CRMcinnes2008enhanced,KNakano2014feasibility,YZhang2014feasibility}. 
Here is assumed that dynamical equation of the SR harvesters are represented altogether by the following formula: 
\begin{eqnarray}
&& m \frac{\dd \dot{x}}{\dd t} = -m \gamma \dot{x} -\frac{\partial V(x, t)}{\partial x} + R(t), \label{eq:SRME} \\
&& -\frac{\partial V(x, t)}{\partial x} = -\frac{\dd U(x)}{\dd x} + h\cos \Omega t \nonumber \\ 
&& \hspace{14.9mm} = -(x^3 -x) + h\cos \Omega t \nonumber 
\end{eqnarray}
where $R(t)$ is given as zero-mean white Gaussian noise of auto-correlation function:  
\begin{eqnarray}
< R(t) R(t + \Delta t) > = 2 \gamma m k T \delta (\Delta t). \label{eq:RR}
\end{eqnarray}
Where $m$ denotes oscillator mass of a harvester, $x$ displacement, $m \gamma$ damping constant, $h\cos \Omega t$ sinusoidal external force, $U(x)$ bi-stable potential, $k$ Bolztman constant, and $T$ noise temperature. 
$\dot{(~)}$ implies time differential $\dd / \dd t$ and $<\,>$ ensemble average operation.
We take dissipative energy of controlled motion at frequency $\Omega$ as harvested energy.

Without loss of generality, we can introduce the previously obtained theoretical relationship from \cite{RTakahashi2005interaction,MIDykman1992phase,LGammaitoni1998stochastic}.
In particular, $Q$ and $\phi$ are defined in \cite{RTakahashi2005interaction} and \cite{MIDykman1992phase}.
\begin{eqnarray}
&& < x > = Q \cos(\Omega t - \phi) \label{eq:x} \\
&&  Q = \frac{h}{kT} \frac{W}{\sqrt{W^2 + \Omega^2 }} \label{eq:Q} \\
&& \phi = \arctan{\frac{(\Omega / \Omega\sub{r})(\Omega\sub{r}^2 W + \Omega^2 kT)}{\Omega\sub{r} W^2 + \Omega^2 kT}} \label{eq:phi} \\
&& \frac{W}{2} = \frac{\omega\sub{0}\omega\sub{b}}{2\pi \gamma m} \exp\left( -\frac{\Delta U}{kT} \right)
\end{eqnarray}
Response of amplitude $Q$ and phase $\phi$ for noise intensity $kT$ are shown in \figref{fig:kT-QPhi}.
$\Omega\sub{r}$ is a relaxation rate defined by ${\dd^2 U / \dd x^2}|_{x=\pm 1}$ \cite{MIDykman1992phase}. 
$W/2$ corresponds to Kramers rate \cite{LGammaitoni1998stochastic} at $\omega_0 = \dd^2 U / \dd x^2|_{x=\pm 1},~ \omega_b = \dd^2 U / \dd x^2|_{x= 0}$, and potential barrier $\Delta U = 1/4$. 
According to the reaction of $Q$ in \figref{fig:kT-QPhi}, SR appears around $kT = 0.1$.
When SR appears, the dissipative energy becomes maximum as confirmed in \cite{CRMcinnes2008enhanced,KNakano2014feasibility,YZhang2014feasibility}.
The results explain that the external force frequency $\Omega$ matches the time lag to overcome the potential barrier for the harvester. 

On the other hand, when a harmonic oscillator has phase lag $\pi / 2$ to sinusoidal external force, so called resonance appears, then dissipative energy becomes maximum.
Similar phenomena is also confirmed in nonlinear oscillators \cite{YSusuki2007ener}.
In \figref{fig:kT-QPhi}, phase lag $\phi$ becomes maximum at SR\footnote{\eqrefs{eq:x}, (\ref{eq:Q}), and (\ref{eq:phi}) are derived for over damped model without inertia, therefore they will not match \eqref{eq:SRME} completely.} \cite{MIDykman1992phase}.
However, it has not been accounted for the relationship between phase lag and dissipative energy. 
\section{Active and reactive power}
\begin{figure}[bt]\centering
 \includegraphics[width=86mm]{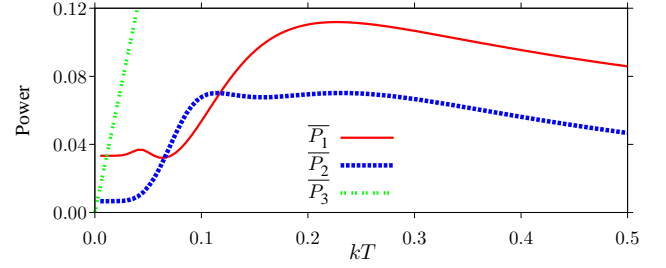}
 \caption{$\overline{P_1}$ (red solid line), $\overline{P_2}$ (blue dotted line), $\overline{P_3}$ (green broken line) as a function of $kT$ calculated theoretically.}
 \label{fig:kT-P}
\end{figure}

Here we focus on the relationship of phase lag $\phi$ and energy flow in system of Eqs. (\ref{eq:SRME}) and (\ref{eq:RR}).
Since there are two external forces; sinusoidal force and noise, it is difficult to decide each contribution.
For energy harvesting by SR, it is necessary to see the energy flow.
From \eqref{eq:SRME}, the following relationship is obtained.
\begin{eqnarray}
 \frac{m}{2} \frac{\dd}{\dd t} <\dot{x}^2> &=& - m\gamma <\dot{x}^2> - < \frac{\dd U}{\dd x} \dot{x} > \nonumber \\ && + < \dot{x} > h\cos \Omega t + < R(t)\dot{x} >. \label{eq:PB}
\end{eqnarray}
Here allocates each term to input, active and reactive power.
Through the mechanical-electrical analogy \cite{KOgata}, electrical voltage and current correspond to externally given mechanical force and the velocity of oscillator, respectively,
so that the correspondence of electrical active power to mechanical dissipative power appears.
At the same time, the reactive power can be explained as the mechanical energy flow. 
Generally active and reactive power are averaged over a period. 
On the other hand, instantaneous input, active, and reactive power are depicted as follows: 
\begin{eqnarray}
&& \mbox{\bf \hspace{-5mm} Input power: } \nonumber \\
&&  < \dot{x} h\cos \Omega t > + < R(t)\dot{x} > \nonumber \\ 
&& \hspace{5mm} = - \frac{1}{\gamma m} \left< \frac{\partial V}{\partial x} \right> h\cos \Omega t + \gamma k T, \label{eq:Pin} \\
&& \mbox{\bf \hspace{-5mm} Active power: }  \nonumber \\
&& < m \gamma \dot{x}^2 > = \gamma k T + \frac{1}{\gamma m} \left< \frac{\partial V}{\partial x} \right>^2, \label{eq:Pac} \\
&& \mbox{\bf \hspace{-5mm} Reactive power: }  \nonumber \\
&& \frac{m}{2}\frac{\dd}{\dd t} <\dot{x}^2> + \left< \frac{\dd U}{\dd x} \dot{x} \right>  \nonumber \\ 
&& \hspace{5mm} = - \frac{1}{\gamma m} \left< \frac{\pp V}{\pp x} \right>^2 - \frac{1}{\gamma m} \left< \frac{\pp V}{\pp x} \right> h\cos \Omega t,  \label{eq:Pre}
\end{eqnarray}
where the following two equations derived from the Fokker-Plank equation are substituted into \eqref{eq:PB}.
\begin{eqnarray}
&& < \dot{x}^2 > = \frac{kT}{m} + \frac{1}{\gamma^2 m^2} \left< \frac{\partial V}{\partial x} \right>^2 \label{eq:dotx2} \\
&& <\dot{x}> = - \frac{1}{\gamma m} \left< \frac{\partial V}{\partial x} \right>. \label{eq:dotx}
\end{eqnarray}
In Eqs.(\ref{eq:Pin}), (\ref{eq:Pac}), and (\ref{eq:Pre}), those power are consisted of following $P\sub{i}~ ({\rm i} = 1,2,3)$: 
\begin{eqnarray}
\hspace{1mm} P_1 = - \frac{1}{\gamma m} \left< \frac{\partial V}{\partial x} \right> h\cos \Omega t, ~~~P_2 = \frac{1}{\gamma m} \left< \frac{\partial V}{\partial x} \right>^2, ~~~P_3 = \gamma k T.   \nonumber   
\end{eqnarray}
Here we average $P\sub{i}~ ({\rm i} = 1,2,3)$ over a period $2\pi / \Omega$, and express them as $\overline{P\sub{i}}~ ({\rm i} = 1,2,3)$.
\figref{fig:kT-P} shows them as functions of noise strength $kT$.
$\overline{P_3}$ is in proportion to $kT$.
$\overline{P_1}$ and $\overline{P_2}$ reach thier peak around SR. 

Here we discuss the physical meaning of $P\sub{i}~ ({\rm i} = 1,2,3)$.
The power sources of $P_1$ and $P_3$ are obvious.
$P_1$ is input power from sinusoidai force $h\cos \Omega t$, 
because $< \dot{x} h\cos \Omega t > = P_1$ can be derived from \eqref{eq:dotx}. 
$P_3$ is input power from noise $R(t)$, 
because $< R(t)\dot{x} > = P_3$ can be derived from Eqs.(\ref{eq:SRME}) and (\ref{eq:RR}).
$P_3$ corresponds to dissipative power consumed by localized small scale vibration. 
On the other hand, $P_2$ can not be derived directly from the given forces, 
and the power source is not clearly defined.
As you can understand, $P_2$ is a term which cancels out by adding both active (\eqref{eq:Pac}) and reactive powers (\eqref{eq:Pre}). 
However, at first, $P_2$ includes frequency $\Omega$, 
since $P_2$ depends on potential shape $V(x,t)$, 
as in Eqs.(\ref{eq:SRME}) and (\ref{eq:x}). 
In addition, $P_2$ becomes maximum around SR as shown in \figref{fig:kT-P}. 
Therefore $P_2$ seems a power to vibrate over the potential barrier periodically. 

Next, energy flow is discussed based on the above power allocation.
The input power $P_3$ from noise goes directly to active power and is consumed. 
\figref{fig:kT-P} and Eqs.(\ref{eq:Pin}), (\ref{eq:Pac}), (\ref{eq:Pre}) leads that, 
when SR appears, $P_1$ contributes to the reactive power mostly.
Then $P_2$ increases, since reactive power ($= -P_2 + P_1$) of \eqref{eq:Pre} has a limit.
Consequently active power $P_2$ including $\Omega$ also increases. 
And $P_2$ equals to the dissipative power of controlled motion.
The special feature is that, when SR appears, phase lag $\phi$ and also $P_2$ are maximized at an appropriate noise intensity.
Hence, it is reasonable to adjust dissipative power by altering $kT$ similar to power factor correction in a circuit.
\section{Numerical estimation}
\begin{figure}[bt]\centering
 \includegraphics[width=87mm]{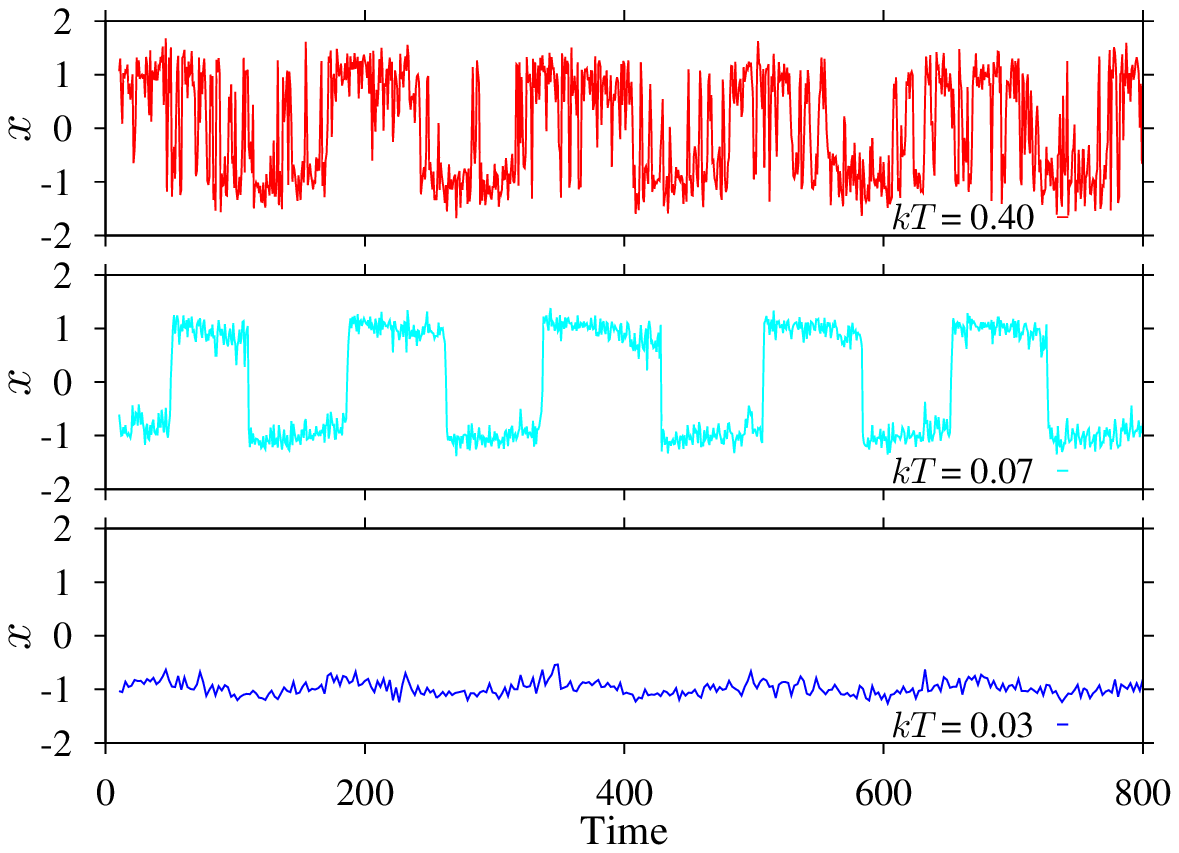}
 \caption{Time changes of displacement $x$ under the noise intensity $kT = 0.03, 0.07, 0.40$ with $h$ and $\Omega$ kept constant.} \label{fig:t-x}
 \includegraphics[width=87mm]{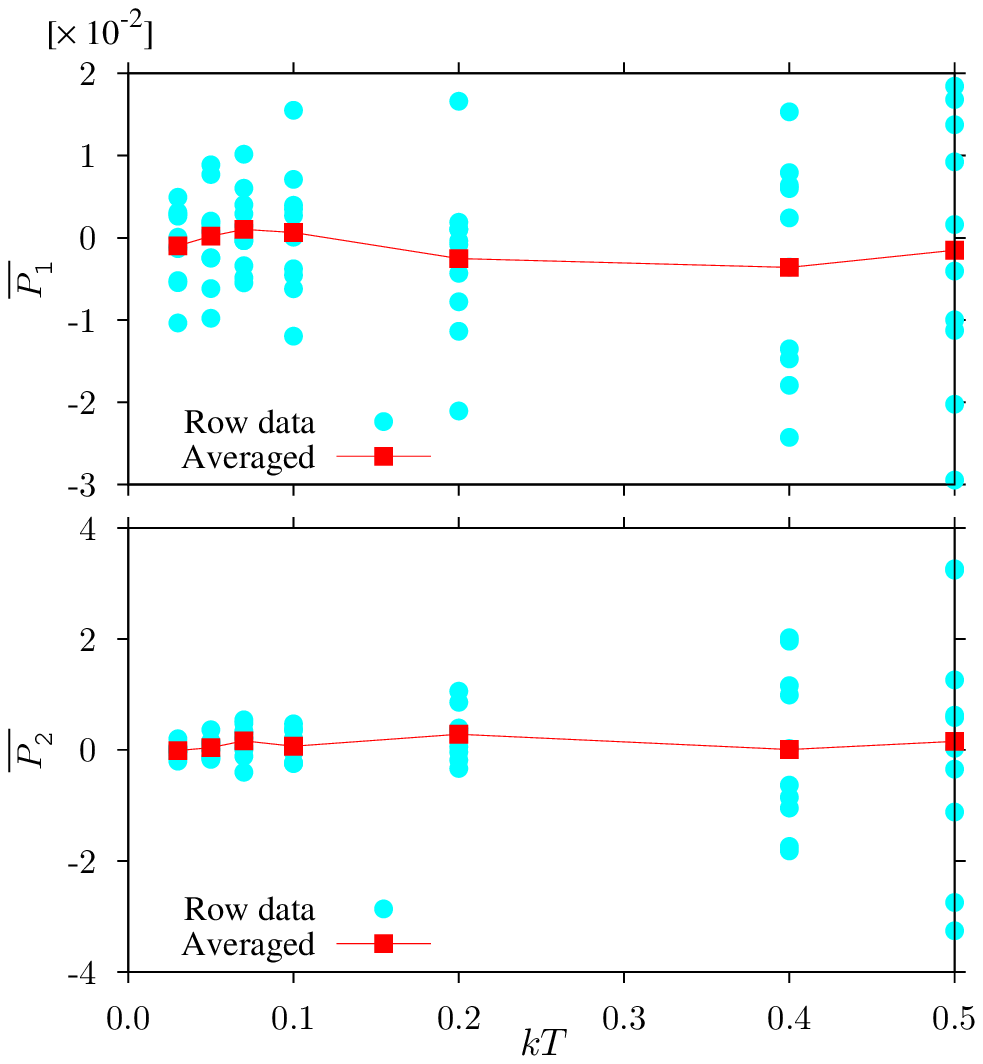}
 \caption{Numerically estimated $\overline{P_1}$ and $\overline{P_2}$ under different noise strength.}
 \label{fig:kT-P_n}
\end{figure}
Figure \ref{fig:t-x} shows time dependence of displacement $x$ at $kT=0.03$, $0.07$, and $0.40$.
SR appears around $kT=0.07$.
\figref{fig:kT-P_n} shows numerically calculated $\overline{P_1}, \overline{P_2}$ from Eqs.(\ref{eq:SRME}) and (\ref{eq:RR}). 
Blue (gray) bullets at each $kT$ are values of ten trials, and red (black) squares are ensenble averaged value from the ten trials.
$\overline{P_1}$ and $\overline{P_2}$ become maximum locally around $kT=0.07$ where SR appears.
However the peaks are not clear as in \figref{fig:kT-P}.
$\overline{P_2}$ increases at $kT=0.5$. 
This is because of a reduction in calculation accuracy as we can see that the variance of blue (gray) bullets increases with noise intensity $kT$ growth.
\section{Conclusion}
In this letter, SR is investigated as a method for energy harvesting from noise.
When SR appears, the phase lag and dissipative power of frequency $\Omega$ are maximized at an appropriate noise intensity.
The result suggests the possibility of maximizing dissipative energy of controlled motion as same as the analogous to power factor correction. \\

\noindent {\bf Acknowledgement} \\ 

\noindent We acknowledge fruitful discussions with Edmon Perkins, visiting researcher supported by NSF-JSPS program.
This work is supported in part by Grant-in-Aid Challenging Exploratory Research No. 26630176. 
MK is financially supported by Kyoto University graduate school of engineering. 


\end{document}